\def\be{\begin{equation}}
\def\ee{\end{equation}}
\def\ba{\begin{array}}
\def\ea{\end{array}}
\documentclass[11pt]{article}
\usepackage{amsfonts}
\usepackage{amssymb}
\usepackage{amsmath,amssymb,amstext,amsfonts,array,hhline,tabularx,graphics}
\usepackage{latexsym}
\usepackage{arydshln}

\begin{document}
\parskip=3pt
\parindent=18pt
\baselineskip=20pt \setcounter{page}{1}

 \hoffset = -1truecm \voffset = -2truecm

 \title{\large\bf Unextendible Maximally Entangled Bases in \(\mathbb {C}^{pd}\otimes \mathbb {C}^{qd}\)}
\date{}
\author{Gui-Jun Zhang$^{1}$, Yuan-Hong Tao$^{1}$ $^\ast$, Yi-Fan Han$^{1}$, Xin-Lei Yong$^{1}$,  Shao-Ming Fei$^{2,3}$\\[10pt]
\footnotesize
\small 1 Department of Mathematics College of Sciences, Yanbian University, Yanji 133002, China\\
\small 2 School of Mathematics Sciences, Capital Normal University, Beijing 100048, China\\
\small 3 Max-Planck-Institute for Mathematics in the Science, 04103 Leipzig, Germany}
\date{}

\maketitle

\centerline{$^\ast$ Correspondence to taoyuanhong12@126.com}
\bigskip

\begin{abstract}
The construction of unextendible maximally entangled bases is tightly related to quantum information
processing like local state discrimination.
We put forward two constructions of UMEBs in \(\mathbb {C}^{pd}\otimes \mathbb {C}^{qd}\)($p\leq q$) based on the constructions of UMEBs in \(\mathbb {C}^{d}\otimes \mathbb {C}^{d}\) and in \(\mathbb {C}^{p}\otimes \mathbb {C}^{q}\), which generalizes
the results in [Phys. Rev. A. 94, 052302 (2016)] by two approaches. Two different 48-member UMEBs in \(\mathbb {C}^{6}\otimes \mathbb {C}^{9}\) have been constructed in detail.
\end{abstract}

\section{Introduction}

\ \ \ \ \ \ It is well known that the quantum states are divided into two parts: separable states and entanglement states. Quantum entanglement, as a potential resource, is widely applied into many quantum information process, such as quantum computation \cite{quco}, quantum teleportation \cite{qutel}, quantum cryptography \cite{qucr} as well as nonlocality \cite{quno}. Nonlocality is a very useful concept in quantum mechanics \cite{quno,quno2,quno3} and is tightly related to entanglement. However, it is proved that the unextendible product bases (UPBs) reveal some nolocality without entanglement \cite{LMSRef6,LMSRef7}. The UPB is a set of incomplete orthogonal product states in bipartite quantum system $\mathbb {C}^{d}\otimes \mathbb {C}^{d^{\prime}}$ consisting of fewer than $dd^{\prime}$ vectors which have no additional product states orthogonal to each element of the set \cite{LMSRef5}.

In 2009, S. Bravyi and J. A. Smolin \cite{LMSRef8} first proposed the notion of unextendible maximally entangled basis(UMEB): a set of incomplete orthogonal maximally entangled states in $\mathbb {C}^{d}\otimes \mathbb {C}^{d^{\prime}}$ consisting of fewer than $dd^{\prime}$ vectors which have no additional maximally entangled vectors that are orthogonal to all of them. These incomplete bases have some special properties. In bipartite space $\mathbb {C}^{d}\otimes \mathbb {C}^{d}$, one can get a state on the UMEB's complementary subspace, whose entanglement of assistance (EoA) is strictly smaller than log$d$, the asymptotic EoA \cite{LMSRef8}. As for in $\mathbb {C}^{d}\otimes \mathbb {C}^{d^{\prime}}$, one can also get a state on the complementary subspace of UMEB, corresponding to a quantum channel, which would not be unital. Besides, it cannot be convex mixtures of unitary operators too \cite{CB}. In addition, for a given mixed state, its Schmidt number is hard to calculate. If we can get a $n$-member UMEB $\{|\phi_i\rangle\}$ in $\mathbb {C}^{d}\otimes \mathbb {C}^{d^{\prime}}$, the Schmidt number of the following state
\[
{\rho}^{\perp}=\frac{1}{dd^{\prime}-n}(I-\sum^n_{i=1}|\phi_i\rangle\langle\phi_i|),
\]
is smaller than $d$ \cite{GY0}. Therefore, different UMEBs can be used to construct different mixed entangled states with limited Schmidt number, even state with different Schmidt number.

In \cite{CB}, B. Chen and S. M. Fei  provided a way to construct UMEBs in some special cases of bipartite system. Then  H. Nan et al. \cite{TYH}, M. S. Li et al. \cite{LMS}, Y. L. Wang et al. \cite{LMS2,LMS3}, Y. Guo \cite{GY}, G. J. Zhang et al. \cite{Z} developed some new results of UMEB in bipartite system. Later, Y.J. Zhang et al. \cite{ZYJ} and Y. Guo et al. \cite{GY2} generalized the notion of UMEB from bipartite systems to multipartite quantum systems. In \cite{GY2}, Y. Guo showed that if there exists an $N-$member UMEB $\{|\psi_j\rangle\}$ in  \(\mathbb {C}^{d}\otimes \mathbb {C}^{d}\), then there exists a ${qd}^2-q(d^2-N)$-member UMEB in \(\mathbb {C}^{qd}\otimes \mathbb {C}^{qd}\) for any $q\in \mathbb{N}$.
Y. Guo et al. \cite{GY3} also proposed the definition of entangled bases with fixed Schmidt number.

In this paper, we study UMEBs in bipartite system  \(\mathbb {C}^{pd}\otimes \mathbb {C}^{qd}\) ($p\leq q$). A systematic way of constructing UMEBs in \(\mathbb {C}^{pd}\otimes \mathbb {C}^{qd}\) from that in \(\mathbb {C}^{d}\otimes \mathbb {C}^{d}\) is presented firstly, and a construction of 48-member UMEB in \(\mathbb {C}^{6}\otimes \mathbb {C}^{9}\) is given as an example. Furthermore, a explicit method to construct UMEBs containing $pqd^2-d(pq-N)$ maximally entangled vectors in \(\mathbb {C}^{pd}\otimes \mathbb {C}^{qd}\) from an $N$-member UMEB in \(\mathbb {C}^{p}\otimes \mathbb {C}^{q}\) is presented. Moreover, another construction of $48$-member UMEB in \(\mathbb {C}^{6}\otimes \mathbb {C}^{9}\) is obtained, thus generalized the results in Yu Guo[phys.Rev.A.94,052302(2016)] by two approaches.

\section{Preliminaries}

\ \ \ \ \ \ Throughout the paper, we denote $[d]^{\prime}=\{0,1,\cdots,d-1\}$ and $[d]^{*}=\{1,2,\cdots,d\}$.

A pure state $|\psi\rangle$ is said to be a maximally entangled state in \(\mathbb {C}^{d}\otimes \mathbb {C}^{d^{\prime}}\)($d\leq d'$) if and only if for a arbitrary given orthonormal basis $\{|i\rangle\}$ of \(\mathbb {C}^{d}\), there exists an orthonormal basis $\{|i^{\prime}\rangle\}$ of \(\mathbb {C}^{d^{\prime}}\) such that $|\psi\rangle$ can be written as $|\psi\rangle=\frac{1}{\sqrt{d}}\sum_{i=0}^{d-1}|i\rangle\otimes|i^{\prime}\rangle$[6].

A set of pure states $\{|\phi_i\rangle\}_{i=0}^{n-1}\in C^d \otimes C^{d'}$ with the following conditions is called an unextendible maximally entangled bases(UMEB) \cite{LMSRef8}:

(i)$|\phi_i\rangle,i\in[n]^{\prime}$ are all maximally entangled states,

(ii)$\langle\phi_i|\phi_j\rangle=\delta_{ij},i,j\in[n]^{\prime}$,

(iii)$n<dd'$, and if a pure state $|\psi\rangle$ meets that $\langle\phi_i|\psi\rangle=0, i\in[n]^{\prime}$, then $|\psi\rangle$ can not be maximally entangled.

Let ${\cal{M}}_{d'\times d}$ be the Hilbert space of all $d'\times  d$ complex matrices equipped with the inner product defined by $\langle A|B \rangle=Tr(A^{\dagger}B)$ for any $A, B\in {\cal{M}}_{d'\times d}$. If $\{A_i\}_{i=0}^{dd'-1}$ constitutes a Hilbert-Schmidt basis of ${\cal{M}}_{d'\times d}$, where $\langle A_i|A_j \rangle =d \delta_{ij}$, then there is a one-to-one correspondence between an orthogonal basis in \(\mathbb {C}^{d}\otimes \mathbb {C}^{d^{\prime}}\) $\{|\phi_i\rangle\}$ and $\{A_i\}$ as follows \cite{GY2,GY3}:
\[
|\phi_i\rangle=\frac{1}{\sqrt{d}}\sum_{k=0}^{d-1}\sum_{\ell'=0}^{d'-1}a_{\ell' k}^{(i)}|k\rangle|\ell'\rangle\in \mathbb {C}^{d}\otimes \mathbb {C}^{d^{\prime}}\ \ \Leftrightarrow \ \ A_i=[a_{\ell' k}^{(i)}] \in {\cal{M}}_{d'\times d},
\]
\begin{equation}\label{1}
Sr(|\phi_i\rangle) = rank(A_i),\ \ \ \  \langle\phi_i|\phi_j\rangle=\frac{1}{d} Tr(A_i^{\dagger}A_j),
\end{equation}
where $Sr(|\phi_i\rangle)$ denotes the Schmidt number of $|\phi_i\rangle$.
Obviously, $|\phi_i\rangle$ is a maximally entangled pure state in $C^d \otimes C^{d'}$ iff $A_i$ is a $d'\times d$ singular-value-1 matrix (a matrix whose singular values all equal to 1). Specially, $A_i$ is a unitary matrix when $d=d^{\prime}$.

For simplicity we adopt the following definitions \cite{GY}.
We call a Hilbert-Schmidt basis $\Omega=\{A_i\}_{i=0}^{d^2-1}$ in ${\cal{M}}_{d\times d}$
a unitary Hilbert-Schmidt basis (UB) of ${\cal{M}}_{d\times d}$ if $A_i$s are unitary matrices, and
a Hilbert-Schmidt basis $\Omega=\{A_i\}_{i=0}^{dd^{\prime}-1}$ in ${\cal{M}}_{d^{\prime}\times d}$
a singular-value-1 Hilbert-Schmidt basis (SV1B) of ${\cal{M}}_{d\times d}$ if $A_i$s are singular-value-1 matrices.
A set of $d\times d$ unitary matrices $\Omega=\{A_i\}_{i=0}^{n-1}$ ($n<d^2$) is called an unextendible unitary Hilbert-Schmidt basis (UUB) of ${\cal{M}}_{d\times d}$ if
(i) Tr($A_i^{\dagger}A_j$)=$d\delta_{ij}$; (ii) if Tr($A_i^{\dagger}X$)=0, $i\in[n]^{\prime}$, then X is not unitary.

{\sf [Definition]} A set of $d\times d^{\prime}$ ($d< d^{\prime}$) singular-value-1 matrices $\Omega=\{A_i\}_{i=1}^{dd^{\prime}}$ ($n<dd^{\prime}$) is called an unextendible singular-value-1 Hilbert-Schmidt basis (USV1B) of ${\cal{M}}_{d\times d}$ if
(a) Tr($A_i^{\dagger}A_j$)=$d\delta_{ij}$;
(b) if Tr($A_i^{\dagger}X$)=0, $i\in[n]^{\prime}$, then X is not a singular-value-1 matrix.

According to the Eq.(1), it is obvious that $\Omega=\{A_i\}_{i=0}^{d^2-1}$ is a UB iff $\{|\phi_i\rangle\}$ is a maximally entangled basis (MEB) of \(\mathbb {C}^{d}\otimes \mathbb {C}^{d}\) while $\Omega=\{A_i\}_{i=0}^{dd^{\prime}-1}$ is a SV1B iff $\{|\phi_i\rangle\}$ is a MEB of \(\mathbb {C}^{d}\otimes \mathbb {C}^{d^{\prime}}\). And $\Omega=\{A_i\}_{i=0}^{n-1}$ ($n<d^2$) is a UUB iff $\{|\phi_i\rangle\}$ is a UMEB of \(\mathbb {C}^{d}\otimes \mathbb {C}^{d}\) while $\Omega=\{A_i\}_{i=0}^{n-1}$ ($n<dd^{\prime}$) is a USV1B iff $\{|\phi_i\rangle\}$ is a UMEB of \(\mathbb {C}^{d}\otimes \mathbb {C}^{d^{\prime}}\)\cite{GY}.

\section{UMEBs in \(\mathbb {C}^{pd}\otimes \mathbb {C}^{qd}\) $(p\leq q)$ from UMEBs in \(\mathbb {C}^{d}\otimes \mathbb {C}^{d}\)}

\ \ \ \ {\bf Theorem 1.} If there is an N-member UMEB $\{|\psi_j\rangle\}$ in  \(\mathbb {C}^{d}\otimes \mathbb {C}^{d}\), then there exists a $pqd^2-p(d^2-N)$-member UMEB in \(\mathbb {C}^{pd}\otimes \mathbb {C}^{qd}\)$(p{\leq}q)$.

{\bf Proof.}
Let $\{W_j=[w_{i^{\prime}i}^j]\}_{j=0}^{N-1}$ be a UUB of ${\cal{M}}_{d\times d}$ corresponding to $\{|\psi_j\rangle\}$,
\[
|\psi_j\rangle=\frac{1}{\sqrt{d}}\sum_{i=0}^{d-1}\sum_{i^{\prime}=0}^{d-1}w_{i^{\prime}i}^j |i\rangle\otimes |i^{\prime}\rangle, \ \ j\in[N]^{\prime}.
\]

Denote
\[
U_{nm}=\sum_{a=0}^{d-1}e^{\frac{2{\pi}na\sqrt{-1}}{d}}|a{\oplus}_dm\rangle{\langle}a|,
\]
\[V_{kl}=\sum_{a=0}^{p-1}e^{\frac{2{\pi}ka\sqrt{-1}}{p}}|a{\oplus}_ql\rangle{\langle}a|,
\]
where $m,n\in[d]^{\prime};\ \ l\in[q]^{\prime};\ \ k\in[p]^{\prime};\ \ j\in[N]^{\prime}$, and
\[
B_{k0}^{j}=V_{k0}{\otimes}W_{j},\ \ k\in[p]^{\prime};\ \ j\in[N]^{\prime},
\]
\[
B_{kl}^{nm}=V_{kl}{\otimes}U_{nm},\ \ k\in[p]^{\prime};l\in[q-1]^{*};m,n\in[d]^{\prime}.
\]
Set $C_1=\{B_{k0}^{j}\}$ and $C_2=\{B_{kl}^{nm}\}$.
Then $C_1{\cup}C_2$ is exactly a USV1B in  ${\cal{M}}_{qd\times pd}$.

Firstly, all $B_{k0}^{j}$ and $B_{kl}^{nm}$ are $qd\times pd$ singular-value-1 matrices, which satisfy the conditions in the Definition:

(a) $Tr[(B_{k0}^{j})^{\dagger}B_{k'0}^{j'}]=pd\delta_{kk'}\delta_{jj'}$, $Tr[(B_{kl}^{nm})^{\dagger}B_{k'l'}^{n'm'}]=pd\delta_{kk'}\delta_{ll'}\delta_{nn'}\delta_{mm'}$ and $Tr[(B_{kl}^{nm})^{\dagger}B_{k'0}^{j}]=0$,
where $j,j'\in[N]^{\prime}$; $k,k'\in[p]^{\prime}$; $l,l'\in[q-1]^{*}$; $n,n',m,m'\in[d]^{\prime}.$

(b) Denote $S$ the matrix space of $(I_{p}, O_{p\times(q-p)})^t\otimes R$, where $t$ stands for matrix transpose, $I_{p}$ is the $p\times p$ identity matrix, $O_{p\times(q-p)}$ is the $p\times(q-p)$ zero matrix and $R\in {\cal{M}}_{d\times d}$. Obviously the dimension of $S^{\bot}$ is $p(q-1)d^2$. Thus $C_2$ is an SV1B of $S^{\bot}$ with $p(q-1)d^2$ elements.

Assume that $D$ is a singular-value-1 matrix in ${\cal{M}}_{qd\times pd}$, which is orthogonal to all matrices in $C_1 \cup C_2$. Since $C_2$ is a SV1B of $S^{\bot}$, then $D\in S$. No loss of generality, set
\[
 D=\left(
\begin{array}{ccccccccccccccc}
A\\
O_1
   \end{array}
 \right),\quad
A=diag(A_1,A_2,\cdots,A_p),\ \ O_1=O_{(q-p)d\times pd},
\]
where $A_h(h\in[p]^{*})$ are all $d\times d$ matrices.
Note that $D$ is orthogonal to each $B_{k0}^j$ in $C_1$, i.e.,
\[
Tr(D^{\dagger}B_{k0}^{j})=0, \ \ k=[p]^{\prime};\quad j\in[N]^{\prime}.
\]
Then
\[
 Tr\left[\left(
\begin{array}{ccccccccccccccc}
A^{\dagger}&O_1^{\dagger}
   \end{array}
 \right)_{pd\times qd}\cdot
 \left(
\begin{array}{ccccccccccccccc}
G\\
O_1
   \end{array}
 \right)_{qd\times pd}\right]=0,
\]
where $G=diag(\omega_p^{0k}W_j,\omega_p^{1k}W_j,\cdots,\omega_p^{(p-1)k}W_j)$, i.e.,
\[
\omega_p^{0k}Tr(A_1^{\dagger}W_j)+\omega_p^{-1k}Tr(A_2^{\dagger}W_j)+\cdots+\omega_p^{(1-p)k}Tr(A_p^{\dagger}W_j)=0.
\]
Hence,
\[
HX_j=0,
\]
where
\[
 H=\left(
\begin{array}{ccccccccccccccc}
1&1&1&\ldots&1\\
1&\omega_p^{p-1}&\omega_p^{p-2}&\ldots&\omega_p^{1}\\
1&\omega_p^{p-2}&\omega_p^{p-4}&\ldots&\omega_p^{2}\\
\vdots&\vdots&\vdots&\ddots&\vdots\\
1&\omega_p^{1}&\omega_p^{2}&\ldots&\omega_p^{p-1}\\
   \end{array}
 \right),\quad
 X_j=\left(\begin{array}{cc}
Tr(A_1^{\dagger}W_j)\\
Tr(A_2^{\dagger}W_j)\\
Tr(A_3^{\dagger}W_j)\\
\vdots\\
Tr(A_p^{\dagger}W_j)
\end{array}
\right).
\]

Obviously, $X_j=O$ for $j\in[N]^{\prime}$ since $detH \neq 0$. That is to say, $Tr(A_h^{\dagger}W_0)=Tr(A_h^{\dagger}W_1)=\cdots=Tr(A_h^{\dagger}W_{N-1})=0$, $h\in[p]^{*}$. As every $A_n$ is orthogonal to each $W_j$, whereas $\{W_j\}$ is a UUB in ${\cal{M}}_{d\times d}$, none of $A_h$ is unitary. Moreover, all the singular values of $A_h$s are also the singular values of $D$. Therefore, $D$ is not a singular-value-1 matrix, which contradicts to the assumption.
Thus, $C_1\cup C_2$ is a USV1B in ${\cal{M}}_{qd\times pd}$. $\Box$

{\bf Example 1.}
A 48-member UMEB in \(\mathbb {C}^{6}\otimes \mathbb {C}^{9}\) from a 6-member UMEB in \(\mathbb {C}^{3}\otimes \mathbb {C}^{3}\).

A 6-member UMEB in \(\mathbb {C}^{3}\otimes \mathbb {C}^{3}\) from Ref.\cite{LMSRef8} is as follows:
\[
W_j=I-(1-e^{\sqrt{-1}\theta})|\psi_j\rangle\langle\psi_j|,  j=[6]^{\prime},
\]
where
\[
|\psi_{0,1}\rangle=\frac{1}{\sqrt{1+\phi^2}}(|0\rangle\pm\phi|1\rangle),
\]
\[
|\psi_{2,3}\rangle=\frac{1}{\sqrt{1+\phi^2}}(|1\rangle\pm\phi|2\rangle),
\]
\[
|\psi_{4,5}\rangle=\frac{1}{\sqrt{1+\phi^2}}(|2\rangle\pm\phi|0\rangle),
\]
with $\phi = (1+\sqrt{5})/2$.

Then, denote
$$
 U_{nm}=
 \left(
\begin{array}{ccccccccccccccc}
0&0&1\\
1&0&0\\
0&1&0\\
   \end{array}
 \right)^m\cdot
 \left(
\begin{array}{ccccccccccccccc}
1&0&0\\
0&\omega_3&0\\
0&0&\omega_3^2\\
   \end{array}
 \right)^n,
 \quad
 m,n\in[3]^{\prime},
 $$

$$
 V_{kl}=
 \left(
\begin{array}{ccccccccccccccc}
0&0&1\\
1&0&0\\
0&1&0\\
   \end{array}
 \right)^l\cdot
 \left(
 \begin{array}{ccccccccccccccc}
1&0\\
0&1\\
0&0\\
   \end{array}
 \right)\cdot
 \left(
\begin{array}{ccccccccccccccc}
1&0\\
0&-1\\
   \end{array}
 \right)^k,
 \quad
 k\in[2]^{\prime};\quad
 l\in[3]^{\prime},
 $$
where $\omega_3=e^{\frac{2\pi\sqrt{-1}}{3}}$.

Let
$$
B_{01}^{nm}=\left(
 \begin{array}{ccccccccccccccc}
0&0\\
U_{nm}&0\\
0&U_{nm}\\
   \end{array}
 \right),
\quad
B_{11}^{nm}=\left(
 \begin{array}{ccccccccccccccc}
0&0\\
U_{nm}&0\\
0&-U_{nm}\\
   \end{array}
 \right),
 \quad
$$
$$
B_{02}^{nm}=\left(
 \begin{array}{ccccccccccccccc}
0&U_{nm}\\
0&0\\
U_{nm}&0\\
   \end{array}
 \right),
 \quad
 B_{12}^{nm}=\left(
 \begin{array}{ccccccccccccccc}
0&U_{nm}\\
0&0\\
-U_{nm}&0\\
   \end{array}
 \right),
 \quad
$$
$$
B_{00}^j=\left(
 \begin{array}{ccccccccccccccc}
W_{j}&0\\
0&W_{j}\\
0&0\\
   \end{array}
 \right),
 \quad
 B_{10}^j=\left(
 \begin{array}{ccccccccccccccc}
W_{j}&0\\
0&-W_{j}\\
0&0\\
   \end{array}
 \right),
 \quad
$$
where $n,m\in[3]^{\prime}$; $j\in[6]^{\prime}$. Set
$C_1=\{B_{k0}^{j}\},\quad C_2=\{B_{kl}^{nm}\}$,
for $k\in[2]^{\prime};\quad j\in[6]^{\prime};\quad l\in[2]^{*};\quad n,m\in[3]^{\prime}$.
According to Theorem 1, we have that $C_1\cup C_2$ is a 48-number UMEB in $\mathbb {C}^{6}\otimes \mathbb {C}^{9}$.

{\it Remark 1.}
Theorem 1 in Ref.\cite{LMS2} is a special case of the above Theorem 1 for $p=q$.

\section{UMEBs in \(\mathbb {C}^{pd}\otimes \mathbb {C}^{qd}\)$(p{\leq}q)$ from UMEBs in \(\mathbb {C}^{p}\otimes \mathbb {C}^{q}\)}

\ \ \ \ \ \ Next, we will present a general approach to construct UMEBs in \(\mathbb {C}^{pd}\otimes \mathbb {C}^{qd}\) from UMEBs in \(\mathbb {C}^{p}\otimes \mathbb {C}^{q}\).

{\bf Theorem 2.}
If there is an N-member UMEB $\{|\psi_j\rangle\}$ in  \(\mathbb {C}^{p}\otimes \mathbb {C}^{q}\), then there exists a $pqd^2-d(pq-N)$-member UMEB in \(\mathbb {C}^{pd}\otimes \mathbb {C}^{qd}\)$(p\leq q)$.

{\bf Proof.}
Let $\{W_j=[w_{i^{\prime}i}^j]\}_{j=0}^{N-1}$ be a USV1B of ${\cal{M}}_{q\times p}$ corresponding to $\{|\psi_j\rangle\}$,
then
\[
|\psi_j\rangle=\frac{1}{\sqrt{d}}\sum_{i=0}^{d-1}\sum_{i^{\prime}=0}^{d^{\prime}-1}w_{i^{\prime}i}^j |i\rangle\otimes |i^{\prime}\rangle, \ \ j\in[N]^{\prime}.
\]

Denote
\[
U_{nm}=\sum_{a=0}^{d-1}e^{\frac{2{\pi}na\sqrt{-1}}{d}}|a{\oplus}_dm\rangle{\langle}a|,
\]
\[
V_{kl}=\sum_{a=0}^{p-1}e^{\frac{2{\pi}ka\sqrt{-1}}{p}}|a{\oplus}_ql\rangle{\langle}a|,
\]
where $m,n\in[d]^{\prime};\ \ l\in[q]^{\prime};\ \ k\in[p]^{\prime};\ \ j\in[N]^{\prime}.$
Let
\[
B_{n0}^{j}=U_{n0}{\otimes}W_{j},\ \ n\in[d]^{\prime};\ \ j\in[N]^{\prime},
\]
\[
B_{nm}^{kl}=U_{nm}{\otimes}V_{kl},\ \ m\in[d-1]^{*};n\in[d]^{\prime};l\in[q]^{\prime};k\in[p]^{\prime},
\]
and
\[
C_1=\{B_{n0}^{j}\},\quad C_2=\{B_{nm}^{kl}\}.
\]
then, $C_1{\cup}C_2$ is exactly a USV1B in ${\cal{M}}_{qd\times pd}$.

Firstly, all $B_{n0}^{j}$ and $B_{nm}^{kl}$ are $qd\times pd$ singular-value-1 matrices, satisfying the conditions in the Definition:

(a) $Tr[(B_{n0}^{j})^{\dagger}B_{n'0}^{j'}]=qd\delta_{nn'}\delta_{jj'}$, $Tr[(B_{nm}^{kl})^{\dagger}B_{n'm'}^{k'l'}]=qd\delta_{nn'}\delta_{mm'}\delta_{kk'}\delta_{ll'}$ and $Tr[(B_{nm}^{kl})^{\dagger}B_{n'0}^{j}]=0,$
where $j,j'\in[N]^{\prime};\  k,k'\in[q]^{\prime};\ l,l'\in[p]^{\prime};\  n,n'\in[d]^{\prime}; \ m,m'\in[d-1]^{*}$.

(b) Denote $S$ the matrix space of $I_d\otimes R$, where $R\in {\cal{M}}_{q\times p}$. Obviously, the dimension of $S^{\bot}$ is $pq(d-1)d$.

Setting $C_1=\{B_{n0}^{j}\}$ and $C_2=\{B_{nm}^{kl}\}$, we have
that $C_2$ with $pq(d-1)d$ elements is an SV1B of $S^{\bot}$, and
$C_1{\cup}C_2$ is just a USV1B in ${\cal{M}}_{qd\times pd}$.

Assume that $D$ is a singular-value-1 matrix in ${\cal{M}}_{qd\times pd}$, which is orthogonal to all matrices in $C_1 \cup C_2$. Since $C_2$ is a SV1B of $S^{\bot}$, then $D\in S$. No loss of generality, set
\[
 D=diag(A_1,A_2,\cdots,A_d)_{qd\times pd},\quad
\]
where $A_h(h\in[d]^{*})$ are all $q\times p$ matrices.
Similar to the proof of Theorem 1, we can prove that none of $A_h$ is singular-value-1 matrices. Moreover, all the singular values of all $A_h$s are also the singular values of $D$, namely, $D$ is not a singular-value-1 matrix, which contradicts to the assumption.
Thus, $C_1\cup C_2$ is a USV1B in ${\cal{M}}_{qd\times pd}$. $\Box$

{\bf Example 2.}\
A 48-member UMEB in \(\mathbb {C}^{6}\otimes \mathbb {C}^{9}\) from a 4-member UMEB in \(\mathbb {C}^{2}\otimes \mathbb {C}^{3}\).

A 4-member UMEB in \(\mathbb {C}^{2}\otimes \mathbb {C}^{3}\) is as follows:
\[
W_{0,1}=|0'\rangle\langle0|\pm|1'\rangle\langle1|,
\]
\[
W_{2,3}=|0'\rangle\langle1|\pm|1'\rangle\langle0|.
\]

Denote
$$
 U_{nm}=
 \left(
\begin{array}{ccccccccccccccc}
0&0&1\\
1&0&0\\
0&1&0\\
   \end{array}
 \right)^m\cdot
 \left(
\begin{array}{ccccccccccccccc}
1&0&0\\
0&\omega_3&0\\
0&0&\omega_3^2\\
   \end{array}
 \right)^n,
 \quad
 m,n\in[3]^{\prime},
 $$

$$
 V_{kl}=
 \left(
\begin{array}{ccccccccccccccc}
0&0&1\\
1&0&0\\
0&1&0\\
   \end{array}
 \right)^l\cdot
 \left(
 \begin{array}{ccccccccccccccc}
1&0\\
0&1\\
0&0\\
   \end{array}
 \right)\cdot
 \left(
\begin{array}{ccccccccccccccc}
1&0\\
0&-1\\
   \end{array}
 \right)^k,
 \quad
 k\in[3]^{\prime};\quad
 l\in[2]^{\prime},
 $$
where $\omega_3=e^{\frac{2\pi\sqrt{-1}}{3}}$.
Let
$$
B_{01}^{kl}=\left(
 \begin{array}{ccccccccccccccc}
0&0&V_{kl}\\
V_{kl}&0&0\\
0&V_{kl}&0\\
   \end{array}
 \right),
B_{11}^{kl}=\left(
 \begin{array}{ccccccccccccccc}
0&0&\omega_3^2V_{kl}\\
V_{kl}&0&0\\
0&\omega_3V_{kl}&0\\
   \end{array}
 \right),
B_{21}^{kl}=\left(
\begin{array}{ccccccccccccccc}
0&0&\omega_3V_{kl}\\
V_{kl}&0&0\\
0&\omega_3^2V_{kl}&0\\
    \end{array}
 \right),
$$
$$
B_{02}^{kl}=\left(
 \begin{array}{ccccccccccccccc}
0&V_{kl}&0\\
0&0&V_{kl}\\
V_{kl}&0&0\\
   \end{array}
 \right),
 B_{12}^{kl}=\left(
 \begin{array}{ccccccccccccccc}
0&\omega_3V_{kl}&0\\
0&0&\omega_3^2V_{kl}\\
V_{kl}&0&0\\
   \end{array}
 \right),
B_{22}^{kl}=\left(
\begin{array}{ccccccccccccccc}
0&\omega_3^2V_{kl}&0\\
0&0&\omega_3V_{kl}\\
V_{kl}&0&0\\
    \end{array}
 \right),
$$
$$
B_{00}^j=\left(
 \begin{array}{ccccccccccccccc}
W_{j}&0&0\\
0&W_{j}&0\\
0&0&W_{j}\\
   \end{array}
 \right),
 B_{10}^j=\left(
 \begin{array}{ccccccccccccccc}
W_{j}&0&0\\
0&\omega_3W_{j}&0\\
0&0&\omega_3^2W_{j}\\
   \end{array}
 \right),
 B_{20}^j=\left(
 \begin{array}{ccccccccccccccc}
W_{j}&0&0\\
0&\omega_3^2W_{j}&0\\
0&0&\omega_3W_{j}\\
   \end{array}
 \right),
$$
where $k\in[3]^{\prime};\  l\in[2]^{\prime};\  j\in[4]^{\prime}$. Set
$C_1=\{B_{n0}^{j}\},\quad C_2=\{B_{nm}^{kl}\},$
for $k\in[3]^{\prime};\  l\in[2]^{\prime};\  j\in[4]^{\prime}; \ n\in[3]^{\prime}; \ m\in[2]^{*}$.
Then according to Theorem 2, $C_1\cup C_2$ is a 48-member UMEB in $\mathbb {C}^{6}\otimes \mathbb {C}^{9}$.

{\bf Remark 2.} \  The Constructions of UMEB in Theorem 1 and Theorem 2 are different, which can be easily seen from the Examples 1 and 2. We can give a state with Schmidt number 4 in the subspace of the UMEB in Example 1. While what we can get in the subspace of the UMEB in Example 2 are the states with Schmidt number no more than 3. In fact, according to Theorem 2, one can construct a UMEB in \(\mathbb {C}^{4}\otimes \mathbb {C}^{6}\) from the UMEB in \(\mathbb {C}^{2}\otimes \mathbb {C}^{3}\), while one can not do this way from the Theorem 1. Here Theorem 1 in \cite{LMS2} is a also special case of the above Theorem 2 for $p=q$.

{\bf Remark 3.} \  By using Theorem 2 in \cite{Z}, we can give a $p(q-r)$-member UMEB in \(\mathbb {C}^{p}\otimes \mathbb {C}^{q}\). According to Theorem 2 in this paper, we can obtain a $pd(qd-r)$-member UMEB in \(\mathbb {C}^{pd}\otimes \mathbb {C}^{qd}\), in whose subspace we can get some states with Schmidt number $dr$. We can also get a $pd(qd-r)$-member UMEB directly by Theorem 2 in \cite{Z}, nevertheless, in the associated subspace, one can only attain the states with Schmidt number no greater than $r$. Therefore, they are different constructions. Actually, there are many $N$-number UMEBs in \(\mathbb {C}^{p}\otimes \mathbb {C}^{q}\), where $p\nmid N$. In this case, it doesn't hold that $pd|(pqd^2-d(pq-N))$. Namely, we can not even get a UMEB with the same number of members by Theorem 2 in \cite{Z}.

\section{Conclusion}
\ \ \ \ \ \ We have provided an explicit way of constructing a $pqd^2-p(d^2-N)$-member UMEB in \(\mathbb {C}^{pd}\otimes \mathbb {C}^{qd}\) from an $N$-member UMEB in \(\mathbb {C}^{d}\otimes \mathbb {C}^{d}\), and constructed a $48$-number UMEB in \(\mathbb {C}^{6}\otimes \mathbb {C}^{9}\) as a detailed example.
We have also established a method to construct a $pqd^2-d(pq-N)$-member UMEB in \(\mathbb {C}^{pd}\otimes \mathbb {C}^{qd}\) from an $N$-member UMEB in \(\mathbb {C}^{p}\otimes \mathbb {C}^{q}\), and presented another $48$-member UMEB in \(\mathbb {C}^{6}\otimes \mathbb {C}^{9}\). These results may highlight the
further investigations on the construction of unextendible bases and the theory of quantum entanglement.

\newpage
\bigskip
\noindent{\sf Acknowledgements}

\noindent The work is supported by the NSFC under number 11675113, 11761073 and NSF of Beijing under No. KZ201810028042.

\bigskip
\noindent{\sf Acknowledgements}

\noindent G.-J.Z and Y.-H.T. wrote the main manuscript text. All of the authors reviewed the manuscript.

\end{document}